\documentclass[twocolumn,prl,aps,showpacs]{revtex4}
\usepackage{epsfig}
\def\be{\begin{equation}}
\def\ee{\end{equation}}

\begin{document}
%%%%%%%%%%%%%%%%%%%%%%%
\title{Landau and Dynamical Instabilities of the Superflow\\ of
Bose-Einstein Condensates in Optical Lattices}
%%%%%%%%%%%%%%%%%%%%%%%
\author{Biao Wu and Qian Niu}
\affiliation{Department of Physics, The University of Texas at Austin,
Austin, Texas 78712-1081}

\date{\today}

\begin{abstract}
\vskip10pt
The superfluidity of Bose-Einstein condensates in optical lattices are 
investigated.   Apart from the usual Landau instability which occurs when a 
BEC flows faster than the speed of sound, the BEC can also suffer a 
dynamical instability, resulting in period doubling and other sorts of 
symmetry breaking of the system.   Such an instability may play a crucial 
role in the dissipative motion of a trapped  BEC in an optical lattice 
recently observed \cite{Burger}.
\end{abstract}

\maketitle

%67.40.Db Quantum statistical theory; ground state, elementary excitations
%03.75.Fi Phase coherent atomic ensembles; quantum condensation phenomena
%05.30.Jp Boson systems
%32.80.Pj Optical cooling of atoms; trapping
%03.75.B  Atom optics

Optical lattices have long been used to manipulate
ultra-cold atoms, with applications ranging from beam splitters
\cite{Gould}, accelerators \cite{Raizen,Dahan} to lithography
\cite{Timp}. There are now growing interests in replacing
the cold atoms with  Bose-Einstein condensates (BECs)
of alkali atoms \cite{Kasevich,Kozuma,Klaus} to explore the effects
of coherence, atomic interaction, and superfluidity, with important
applications in atom lasers \cite{Mewes} and high-precision
interferometry \cite{Carusotto}.

In this Letter, we investigate the superfluidity and instabilities of BECs
in optical lattices. In free space, the superflow of a uniform BEC is 
represented by a plane wave,
which has Landau instability when it travels faster than the sound.
In an optical lattice, the natural objects of concern are the BEC Bloch 
waves,
whose amplitudes are modulated with the same periodicity as the lattice.
In the central region of the Brillouin zone of the lowest band,  they are
found to be local energy minima against all sorts of perturbations and thus
represent the superflows of the BEC in optical lattices.
For sufficiently strong repulsive interactions between the atoms,  the 
superfluidity
region can extend over the entire Brillouin zone; but for weaker
interaction, Landau instability can occur
in the outer regions of the zone.
Many of the Bloch states with Landau instability can even be dynamically 
unstable
in that small initial disturbances
around them grow exponentially in time, resulting in period
doubling and other forms of spontaneous breaking of
the periodicity of the system.  This dynamical instability is
unique to BEC Bloch waves and not present
in BEC plane waves in free space.
We map out the dangerous zones of the dynamical instability,
characterize the growth rates, and discuss the experimental
consequences.

We consider the situation of a one dimensional optical lattice
in which the motion in the perpendicular directions are confined
\cite{Burger} or can be disregarded \cite{Klaus}.   We treat the
atomic interaction with the mean field theory, and obtain the
grand canonical Hamiltonian
\be\label{eq:h}
H=\!\int_{-\infty}^{\infty}\!{\rm d}x~\{\phi^*(-{1\over 2}
{\partial^2\over \partial x^2}+v\cos{x})\phi+{c\over 2}|\phi|^4-\mu|\phi|^2\},
\ee
where all the variables are scaled to be dimensionless by the system's
basic parameters, the atomic mass $m$,
the wave number $k_L$ of the two laser lights that generate the optical
lattice, and the average density $n_0$ of the BEC. The chemical potential
$\mu$ and
the strength of the periodic potential $v$ are in units of
${4\hbar^2 k_L^2\over m}$, the wave function $\phi$ in units of $\sqrt{n_0}$,
$x$ in units of ${1\over 2k_L}$, and $t$ in units of
${m\over 4\hbar k_L^2}$. The coupling constant $c={\pi n_0 a_s\over k_L^2}$,
where $a_s>0$ is the $s$-wave scattering length.

Hamiltonian (\ref{eq:h}) is extremized by states in the form of Bloch waves,
$\phi_k(x)=e^{ikx}\varphi_k(x)$, where $\varphi_k(x)$ is of the
period of the optical lattice and can be expanded as a Fourier
series. To find the numerical solution of
a Bloch state in the lowest band, $\varphi_k(x)$,
we truncate the series up to $N$th term 
$\varphi_k(x)=\sum_{-N}^N a_m e^{imx}\,$ (we used $N=10$).
The numerical solution is obtained by varying $\{a_m\}$ so that
the wave function $\varphi_k(x)$ minimizes the system's total energy.
The accuracy is checked by substituting
the solutions into the Gross-Pitaevskii equation,
$
-{1\over 2}{\partial^2\over \partial x^2}\phi+v\cos x\,\phi
+c|\phi|^2\phi=\mu\phi\,,
$
which is obtained  by the variation of Hamiltonian (\ref{eq:h}).

To determine the superfluidity of these Bloch states, we need to find out if
they remain energy minima against perturbations which break the periodicity.
These perturbations can be decomposed  into different modes labeled by $q$,
\be \label{eq:pert}
\delta\varphi_k(x,q) = u_k(x,q)e^{iqx}+v_{k}^*(x,q)e^{-iqx},
\ee
where $q$ ranges between $-1/2$ and $1/2$ and the perturbation functions
$u_k$ and $v_k$ are of periodicity of $2\pi$ in $x$.
Since the system is periodic, the quadratic form of the energy deviation
from the Bloch state $\phi_k$ is block diagonal in $q$, with
each block given by
\be
\delta E_k=\int_{-\infty}^{\infty}{\rm d}x
\pmatrix{u_{k}^*,v_{k}^*}M_k(q)\pmatrix{u_{k} \cr v_{k}},
\label{eq:approx}
\ee
where
\be
M_k(q)=\pmatrix{{\mathcal L}(k+q) & c\varphi_k^2
\cr c\varphi_k^{*2} & {\mathcal L}(-k+q)},
\ee
with
\be
{\mathcal L}(k)=-{1\over 2}({\partial \over \partial x} + ik)^2
+v\cos x-\mu+2c|\varphi_k|^2 .
\ee
If $M_k(q)$ is positive definite for all $-1/2\le q\le 1/2$, the Bloch wave
$\phi_k$ is a local minimum. Otherwise, $\delta E_k$ can be
negative for some $q$, and the Bloch wave is a saddle point.

We first consider the special case $v=0$, BEC in free space, where the
Bloch state $\phi_k$ becomes a plane wave $e^{ikx}$.
The operator $M_k(q)$ becomes a $2\times 2$ matrix
\be
M_k(q)=\pmatrix{q^2/2+kq+c&c\cr c&q^2/2-kq+c}
\ee
whose eigenvalues are found easily as
\be\label{eq:lambda}
\lambda_{\pm}={q^2\over 2}+c\pm\sqrt{k^2q^2+c^2}.
\ee
Since $\lambda_{+}$ is always positive, $M_k(q)$ fails to be
positive definite only when $\lambda_{-}\le 0$, or equivalently,
$|k|\ge \sqrt{q^2/4+c}$. It immediately follows that
the BEC flow $e^{ikx}$ becomes a saddle point when the flow speed exceeds
the sound speed, $|k|>\sqrt{c}$.  This is exactly the Landau condition
for the breakdown of superfluidity \cite{Lifshitz}, which has recently
been confirmed experimentally on BEC \cite{Raman}.

The stability phase diagrams for BEC Bloch waves are shown in the panels
of Fig.1, where different values of $v$ and $c$ are considered.
The results have reflection symmetry in $k$ and $q$, so we only
show the parameter region, $0 \le k \le 1/2$ and
$0 \le q \le 1/2$.  In the shaded area (light or dark) of each panel,
the matrix $M_k(q)$ has negative eigenvalues, and the corresponding Bloch
states $\phi_k$ are saddle points and 
have Landau instability. For those values of $k$ outside the shaded
area, the Bloch states are local energy minima and represent 
super-flows.   The
super-flow region expands with increasing atomic interaction $c$, and
occupies the entire Brillouin zone for sufficiently large $c$.
On the other hand, the lattice potential strength $v$ does not affect
the super-flow region very much as we see in each row.
The phase boundaries for $v\ll 1$ are well reproduced from the analytical
expression $k=\sqrt{q^2/4+c}$ for $v=0$, which is plotted as triangles
in the first column.

\begin{figure}[!htb]
\begin{center}
\resizebox *{8.5cm}{12cm}{\includegraphics*{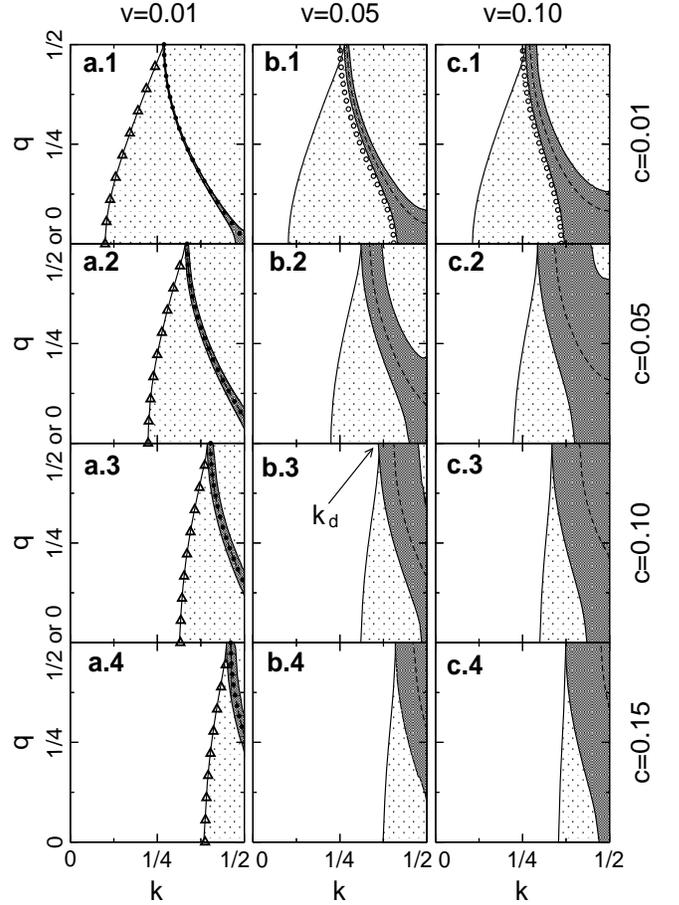}}
\end{center}
\caption{Stability phase diagram of BEC Bloch states in optical
lattices.  $k$ is the wave number of BEC Bloch states;
$q$ denotes the wave number of perturbation modes.
In the shaded (light or dark) area, the perturbation mode has negative
excitation energy; in the dark shaded area, the mode grows or decays
exponentially in time.  The triangles in (a.1-a.4) represent
the boundary, $q^2/4+c=k^2$, of saddle point regions at $v=0$.  The solid
dots in
the first column are from the analytical results of Eq.(\ref{eq:kqc}).
The circles in (b.1) and (c.1) are based on the analytical expression
(\ref{eq:bb}).
The dashed lines indicate the most unstable modes for each Bloch state $k$.}
\label{fig:kq}
\end{figure}

A saddle-point Bloch state $\phi_k$ can still be dynamically stable in that
small deviations from it remain small in the course of time evolution if no
external persistent perturbations are present.  This is the case for all 
Bloch
states either in the absence of atomic interactions or periodic potentials.
When both factors are present, many of the saddle-point Bloch-states become
dynamically unstable against certain perturbation modes $q$, shown as the
dark-shaded regions in Fig.\,\ref{fig:kq}.
These results are obtained from the linear
stability analysis of the Gross-Pitaevskii equation \cite{Gross}, 
\be
i{\partial \over \partial t}\phi=
-{1\over 2}{\partial^2\phi\over \partial x^2}
+v\cos x~\phi+c|\phi|^2\phi\,,
\label{eq:motion}
\ee
which governs the dynamics of the system.
A Bloch state $\phi_k$ is a stationary solution of this equation, depending
on time only through the phase factor $e^{-i\mu t}$.  Writing the deviation
in the similar form of Eq.(\ref{eq:pert}), and expanding the above equation
to first order, we find
\be\label{eq:first}
i{\partial\over \partial t}\pmatrix{u_{k} \cr v_{k}}=\sigma M_k(q)
\pmatrix{u_{k} \cr v_{k}},~~~\sigma=\pmatrix{I&0\cr0&-I}.
\ee
The dynamical stability of the Bloch state $\phi_k$ is determined by
the eigenvalues $\varepsilon_k(q)$ of the matrix $\sigma M_k(q)$.
If they are real for all $-1/2 \le q \le  1/2$,
the state is dynamically stable; otherwise it is dynamically
unstable.

Before discussing our detailed results on the dynamical instability, we pause
here to make some general remarks.
({\bf i}) When all the eigenvalues $\varepsilon_k(q)$
are real, the motions around the Bloch state $\phi_k$ are
oscillations, which can be quantized to yield the phonon excitations
\cite{Lifshitz,Fetter}. The traditional Bogoliubov approach
yields the same matrix, $\sigma M_k(q)$, for the phonon spectrum.
However, the bosonic commutation relation for the phonon operators
imposes the skewed normalization condition $X^{\dagger}\sigma X=1$, which
selects only half of all the modes.  The other half, satisfying
$X^{\dagger}\sigma X=-1$, will be called antiphonon modes for
ease of reference, but they really do not represent physical degrees of 
freedom
independent of the phonons.
({\bf ii}) When the Bloch state is a local minimum ($M_k(q)$ positive
definite),
the dynamical eigenvalues $\varepsilon_k(q)$ of $\sigma M_k(q)$ are all real
and the phonon branch of the spectrum is positive.  The key to the proof is
to notice
that
\be\label{eq:proof}
\varepsilon_k(q) X^{\dagger}\sigma X= X^{\dagger}M_k(q)X
\ee
for an eigenvector $X$ of $\sigma M_k(q)$. Because
the right hand side is positive and $X^{\dagger}\sigma X$ is real,
$\varepsilon_k(q)$ is real and has the same sign as $X^{\dagger}\sigma X$.
The physical meaning of this theorem is that, when it is a local minimum,
the Bloch state $\phi_k$ is dynamically stable and its phonon
excitations are not energetically favored.
({\bf iii}) Because the matrix $\sigma M_k(q)$ is non-hermitian and
real (when expressed in the momentum representation),
complex eigenvalues can only appear in conjugate pairs, corresponding to 
modes
growing or decaying exponentially at rates given by the imaginary part of the
eigenvalues.  Because both the quadratic forms in Eq.(\ref{eq:proof}) are 
real,
they must vanish when $\varepsilon_k(q)$ is complex. It is then impossible to
enforce the normalization condition $X^{\dagger}\sigma X=1$, corresponding to
the fact that such modes cannot be quantized.

We now present our detailed results on the dynamical stability.
Again, we first look at the case $v=0$, where the eigenvalues of
$\sigma M_k(q)$ are
\be\label{eq:epsilon}
\varepsilon_{\pm}(q)=kq\pm\sqrt{q^2c+q^4/4}.
\ee
These eigenvalues are always real; the BEC flows in free
space are always dynamically stable.
When $v\neq 0$, the situation is totally different: the eigenvalues
$\varepsilon_k(q)$ of $\sigma M_k(q)$ can be complex and Bloch
states can be dynamically unstable. The dark-shaded areas in
Fig.\,\ref{fig:kq} are the places where these $\varepsilon_k(q)$ are complex.

In the first column of Fig.\,\ref{fig:kq}, where $v\ll 1$, the
dark-shaded areas are like broadened curves.
These curves are the solutions of
$\varepsilon_{+}(q-1)=\varepsilon_{-}(q)$,
\be \label{eq:kqc}
k=\sqrt{q^2 c+q^4/4}+\sqrt{(q-1)^2 c+(q-1)^4/4},
\ee
which are plotted as solid dots in Fig.\,\ref{fig:kq}.
This is the resonant condition for a phonon mode to couple with
an antiphonon mode by first order Bragg scattering.
The resonance is necessary because the complex
eigenvalues can appear only in pairs, and they must come from a
pair of real degenerated eigenvalues.  Resonances within the
phonon spectrum or within the antiphonon spectrum do not give rise to
dynamical instability; they only generate gaps in the spectra.
Somehow, in order to produce a mode with zero normalization, one must couple
a pair of modes with opposite normalizations.

In the first row of Fig.\,\ref{fig:kq}, we have another
extreme case $c\ll v$. The open circles along
the left edges of these two dark-shaded areas are given by
\be\label{eq:bb}
E_1(k+q)-E_1(k)=E_1(k)-E_1(k-q)
\ee
where  $E_1(k)$ is the lowest Bloch band of
$H_0=-{1\over 2}{\partial^2\over \partial x^2}+v\cos x$.
When $c=0$, this periodic system is linear; the excitation spectrum
just corresponds to transitions from the the condensate of energy $E_1(k)$
to other Bloch states of energy $E_n(k+q)$,  or vise versa.
The above equation is just the resonant condition between such excitations in
the lowest band ($n=1$).  Alternatively, this condition may be viewed as
the energy and
momentum conservation for two particles in the condensate
to interact and decay into two different Bloch states $E_1(k+q)$ and
$E_1(k-q)$.

One common feature of all the diagrams in Fig.\,\ref{fig:kq} is that
there is a critical Bloch wave number $k_d$ beyond which
the Bloch states $\phi_k$ are dynamically unstable.
The onset instability at $k_d$ always corresponds to $q=1/2$.
In other words, if we drive the Bloch state $\phi_k$ from $k=0$ to $k=1/2$
the first unstable mode appearing is always $q=\pm 1/2$, which
represents period doubling.  Only for $kk_d$ can longer wavelength
instabilities
occur. The growth of these unstable modes drives the system
far away from the Bloch state and spontaneously breaks the translation
symmetry of
the system.  The critical value of the Bloch wave number for the case of
$v\ll 1$ is found to be $k_d=(c+1/16)^{1\over 2}$ by substituting $q={1\over
2}$ into Eq.(\ref{eq:kqc}). In the other extreme case, $c\ll v$,
the same substitution in Eq.(\ref{eq:bb}) yields $k_d=1/4$ with
the help of periodicity of the band energy.  Based on these results and the
diagrams in  Fig.\,\ref{fig:kq}, we find that $k_d \ge {1\over 4}$.

\begin{figure}[!htb]
\begin{center}
\resizebox *{8.5cm}{8.5cm}{\includegraphics*{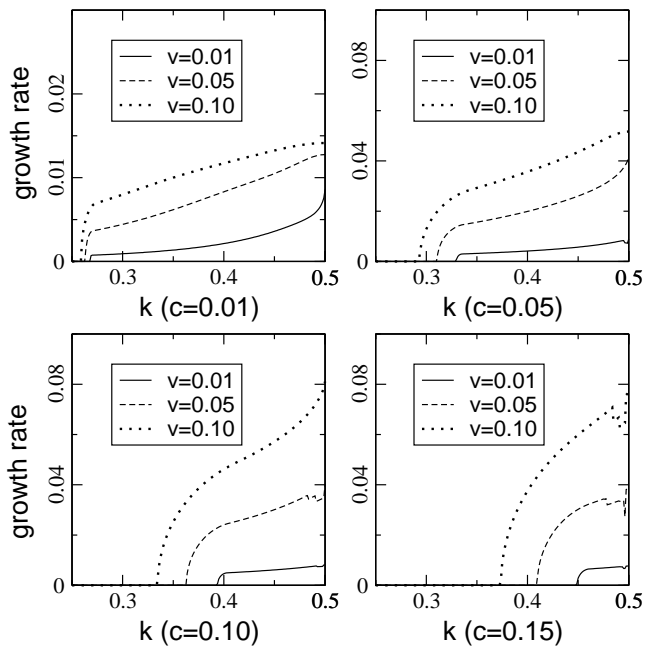}}
\end{center}
\caption{Growth rates of the most unstable modes
of each Bloch state $\phi_k$.
The erratic behavior of some curves
around $k=1/2$ is due to the difficulty in finding the accurate Bloch
waves $\phi_k$ in this region when $c>v$ \cite{Wu}.}
\label{fig:max}
\end{figure}

%%%%%%%%%%%%%%%%%%%%%%%%%%%%%%%%%%%%%%%%%%%%%%%%%%%%%%%%%%%%%%%%%
The dynamical instability discovered in this work should be observable
in experiments. We have mapped out the dangerous zones of dynamical
instability, which give us a good sense of where to look for unstable
Bloch states and modes of instability. In Fig.\,\ref{fig:max},
the rate of growth  $r$ for the most prominent mode
(dashed lines in Fig.\,\ref{fig:kq}) of each Bloch state $k$ is plotted
in Fig.\,\ref{fig:max}.
The physical unit of the growth rate is ${4\hbar k_L^2\over m}$, which is
4.0 per microsecond for sodium and 0.16 per microsecond for rubidium.
Since the lifetime of BECs can be up to the order of
seconds \cite{Kurn}, these growth rates in Fig.\,\ref{fig:max} are
significant.  It is possible to directly observe the change of periodicity 
of the BEC due to the dynamical instability by monitoring the Bragg 
scattering of a probing laser light by the BEC cloud \cite{Phillips}.
This dynamical instability can also cause the disruption of Bloch
oscillations \cite{Wu}.

In a recent experiment, the superfluidity and instabilities of a BEC in an 
optical lattice
was studied using a cigar-shaped (one dimensional) magnetic trap 
\cite{Burger}.
After the BEC was prepared in the trap in the presence of the optical lattice,
the trap was suddenly shifted by $\Delta x$ along the longitudinal direction.
This is equivalent to displacing the whole BEC off the center of the 
harmonic trap then
releasing it.  The subsequent oscillations of the BEC are non-damped if the 
initial displacement
is small, but become dissipative if $\Delta x$ is over a critical value 
$\Delta x_c$.
This qualitatively agrees with our stability diagrams, because larger 
$\Delta x$ implies larger
velocity and therefore larger $k$.    The dissipative behavior was 
explained as a manifestation
of the Landau instability, but the dynamical instability discussed here is 
likely to play a crucial role in our view.  First,  the experiment has $v\sim 0.2$ and 
$c\sim 0.02$, where
the dynamical instability is rampant according to Fig.2.   Second,  $\Delta 
x_c$ increases
with decreasing lattice potential $v$, which is in accordance with the 
trend of the growth rate
as a function of $v$ and $k$ shown in the figure.   Third,  there is no 
dissipation when
the BEC density is low, where Landau instability should be very strong but 
dynamical
instability be very weak according to our Fig.1.   However, more detailed 
analyses are
needed to take into account of the effects of inhomogeneity and thermal 
cloud before
a quantitative comparison with the experiment.    

We are grateful for helpful discussions with Mark Raizen and
Roberto Diener, and supports by the NSF,
the Robert A. Welch Foundation, and the NSF of China. \\
\hrule~\\

\end{document}